\documentclass{bulsao}
\usepackage{graphicx}
\usepackage{latexsym}

\begin{document}

\title{Formation of Ionization-Cone Structures in Active Galactic Nuclei:
II. Nonlinear Hydrodynamic Modelling }

\author{V.~L. Afanasiev$^1$, S.~N. Dodonov$^1$, S.~S. Khrapov$^2$, V.~V. Mustsevoi$^2$, A.~V.
Moiseev$^1$}

\institute{Special Astrophysical Observatory, RAS, Nizhnii Arkhyz,
Karachai-Cherkessian Republic, 357147 Russia \and Volgograd State University,
Volgograd, 400062 Russia}

\offprints{A.V. Moiseev, \email{moisav@sao.ru}}

\date{received:September 21, 2006/revised: November 24, 2006}

\titlerunning{Formation of Ionization-Cone Structures.. II.}

\authorrunning{Afanasiev  et al. }

\abstract{ In Part I of this paper  we described an equilibrium model of a
jet in the gravitational field corresponding to the rigid-rotation region of
the galactic disk. We used linear stability analysis to find the
waveguide-resonance instability of internal gravity waves due to the
superreflection of these waves from the jet boundary. In this part of the
paper, we perform nonlinear numerical 2D and 3D simulations of the
development of this instability. We show that the shocks produced by this
instability in the ambient medium of the jet are localized inside a cone
with a large opening angle and are capable of producing features that are
morphologically similar to those observed in galaxies with active nuclei
(NGC\,5252 for example).}


\maketitle

\section{Introduction}

In Paper I (\cite{paper1}) we used linear stability analysis to investigate
the possibility of the formation of regular structures resulting from the
development of hydrodynamic instabilities in conical jets including those
with large opening angles. At the same time, the very existence of such
weakly collimated jet outflows from active galactic nuclei raises certain
doubts --- observed formations are widely believed to be radiation cones and
not mass outflows (see the ``Introduction'' section in Paper I). From purely
hydrodynamic viewpoint it is not quite clear why in such similar
accretion-jet objects protostellar systems develop highly collimated jets,
whereas active galactic nuclei produce wide-angle cone outflows. Note that
it is extremely difficult to determine from observations whether we are
dealing with a highly collimated or large-opening-angle conical outflow.
This is due to the fact that jets from active galactic nuclei are almost
without exception observed through high-intensity shocks produced by the
intrusion of the outflow matter into the unperturbed ambient medium
(\cite{fal1}, \cite{nag}). Hence a significant number of shock-excited ions
is present along the line of sight in any case.

In this paper we analyze a situation that is alternative to that
addressed in Paper I. In our case, the jet outflow from the
nucleus is collimated, but the nonlinear stage of the development
of instability in the jet results in the localization of the
resulting shocks inside a wide cone. We believe that this effect
can explain the formation of structures observed in the vicinity
of Seyfert nuclei (Z--shaped ``arms'', ``arcs'', and ``arches''
--- see Paper I). Note that Hardee (1982), Falcke et al. (1996)
  and Lobanov et al. (2006)  already pointed out the
possibility of the formation of such structures in the vicinity of
active nuclei due to the interaction of radio-jet matter with
the ambient medium in the boundary layer.

Section \ref{2} describes the technique of numerical simulations,
 Section \ref{3} discusses the results of these simulations and
proves that perturbations that build up in the jet as a result of
waveguide-resonance instability produce in the ambient medium a system of
intensely radiating nonlinear  waves. Section \ref{4} summarizes the main
conclusions and gives the final comments.

\section{Technique of numerical simulations}
\label{2}

The stationary model that we use in this part of the work is quite
similar to that described in Paper I. The only exceptions are the
numerical parameter values that we give at the end of this
section.

\subsection{Basic Equations}

In our numerical simulations of the dynamics of perturbations we
use the following set of hydrodynamics equations in divergent
form written in the spherical coordinate system ($r, \theta,
\varphi$):
\begin{equation}
\frac{\partial \rho }{ \partial t} + {\rm div}(\rho {\bf V}) = 0\,,
\label{e1}
\end{equation}

\begin{equation}
\frac{\partial (\rho\, U)}{  \partial t} + {\rm div}(\rho \,U {\bf V}) =
-\frac{\partial p }{ \partial r} + \rho\, \left(\frac{|{\bf V}|^2 }{ r } -
\frac{\partial \Psi}{  \partial r}\right)\,, \label{e2}
\end{equation}

\begin{equation}
\frac {\partial (r \rho\, W)}{ \partial t} + {\rm div}(r\rho \,W {\bf  V}) =
- \frac{\partial p}{  \partial \theta} + \rho V^2 \mbox{ctg} \theta \,,
\label{e3}
\end{equation}

\begin{equation}
\frac{\partial (r \sin \theta \rho\, V)}{  \partial t} + {\rm div}(r \sin
\theta \rho \,V {\bf  V}) = - \frac{\partial p }{ \partial \varphi} \,,
\label{e4}
\end{equation}

\begin{equation}
\begin{array}{l}
\frac {\partial E }{ \partial t} + {\rm div}\left[{\bf  V} (E+p)\right] =
\\ \qquad\qquad\displaystyle
- \rho\, U\, \frac{\partial \Psi}{ \partial r} + C_\Gamma \rho \,\varepsilon
- C_\Lambda \rho^2 \, \varepsilon^{5/2}\,, \qquad \label{e5}
\end{array}
\end{equation}
where
{\small
$$
{\rm div} {\bf  V} = \frac{1}{  r^2} \frac{\partial}{  \partial r}
(r^2 \, U) + \frac{1 }{ r \sin \theta} \frac{\partial}{  \partial
\theta} (\sin \theta \, W) + \frac{1 }{ r \sin \theta} \frac{\partial V}
 {\partial \varphi},
$$
}
\noindent $\rho$ is the density; $p$, the pressure; ${\bf  V} = (U, W, V)$,
the velocity vector; $\displaystyle E = \rho \frac {|{\bf  V}|^2 }
{2} + \frac{p}{ (\gamma - 1)}$, the total energy, and $\displaystyle
\varepsilon = \frac{p}{  (\gamma - 1) \rho}$, the internal energy
per unit mass.

\subsection{Numerical Scheme and Boundary Conditions}

We numerically integrate equation set (\ref{e1})--(\ref{e5}) using the
\mbox{TVD--E} scheme (\cite{ryu}) adapted to the spherical coordinate system
with a variable radial step.

To analyze the dynamics of axisymmetric perturbations, we use a
two-dimensional (2D) scheme ($r,\theta$) with integration domain
($r_{in} \le r \le r_{ex}$ and $0\le \theta \le \pi/2$) containing a
total of $N_r \times N_\theta$ cells. To analyze the dynamics of
nonaxisymmetric perturbations, we use a three-dimensional (3D)
scheme ($r,\theta,\varphi$) with integration domain ($r_{in} \le r
\le r_{ex}$; $0\le \theta \le \pi/2$; $0\le \varphi \le 2\pi$)
containing a total of $N_r \times N_\theta \times N_\varphi$
cells.

It follows from linear analysis (Paper~I) that the perturbation
wavelength increases linearly with distance, because $kr = {\rm
const}$. Accordingly, in our numerical simulations we use
a variable step in radial coordinate $\Delta r_{i+1} = e^{\Delta x}
\Delta r_i$, where $\Delta x = (\ln r_{ex} - \ln r_{in})/N_r$.
With $\Delta r$ so defined, the number of cells per wavelength
remains constant along the radius. We set constant integration
steps in the $\theta$ and $\varphi$ directions, i.e., $\Delta
\theta = \pi/2N_\theta$, $\Delta \varphi = 2\pi/N_\varphi$.

To avoid distorsions in the ($r, \theta$) plane due to the
differences between the scheme velocities of the propagation of
perturbations in the $r$ and $\theta$ directions,  $\Delta r_i$
and $r_i \Delta \theta$ cells should have equal lengths in these
directions. I.e., $\Delta x = \Delta \theta$ and $N_r = 2N_\theta\,
(\ln r_{ex} - \ln r_{in})/\pi$.

For unstable modes to build up in numerical simulations, a
finite-width transition layer is required between the jet matter
and the ambient medium (see Paper I). We set the extent of the
transition layer by setting the size of the cell in the
$\theta$--direction. In this case, the thickness of the transition
layer decreases with increasing $N_\theta$. The equilibrium
distributions in the transition layer have the following form:
$$
U_s(r) = U_j(r)/2\,,
$$
$$
\displaystyle \rho_s(r) = \frac{\tilde{R} \rho_j(r) r^2 }{ \Omega^2 r^2 +
U_j^2(r)/4}\,,
$$
$$
p_s(r) = p_j(r) = p_a(r)\,,
$$
where subscripts ``j'', ``a'', and ``s'' refer to the jet, ambient
medium, and transition layer, respectively, and $U_j$ and $U_s$
are the radial velocities of gas flow in the equilibrium state.

We set the initial perturbation of the $\theta$ component in the
following form:
$$
\begin{array}{l}
\tilde W(r,\theta, \varphi) = \Delta_w \, \sin(kr\,\ln r + m\,
\varphi)\, 
 \\
\qquad\qquad \displaystyle
\times \exp\left\{-\left({\frac{\sin \theta - sin \theta_j}
{\delta}}\right)^2 \right\},
\end{array}
$$
where $\Delta_w$ is the initial perturbation amplitude ($\Delta_w
\ll U_j$); $kr$, the dimensionless wavenumber, and  \mbox{$m$},
the number of azimuth mode. We set the scale factor $\delta$ equal
to 0.2.

We use the following boundary condition.

\begin{itemize}
\item In the symmetry plane of the system ($\theta = \pi / 2$):
symmetric boundary conditions  $f(\pi/2-0)=f(\pi/2+0)$ for $E$,
$\rho$, $U$, and $V$, and antisymmetric boundary conditions
$f(\pi/2-0)=-f(\pi/2+0)$ for $W$.
\item  At the symmetry axis of the system ($\theta$
= 0):  2D scheme --- symmetric boundary conditions $f(-0)=f(+0)$
for $E$, $\rho$, and $U$ and antisymmetric boundary conditions
$f(-0)=-f(+0)$ for $W$; 3D-scheme --- $f(r,-0,\varphi) = f(r, +0
, \varphi + \pi)$.
\item At $\varphi = 0$ and $\varphi = 2\pi$: periodic boundary
conditions --- $f(-0)=f(2\pi-0)$, $f(2\pi+0)=f(+0)$.
\item At the inner ($r = r_{in}$) and outer ($r = r_{ex}$)
radial boundaries:\\
{\footnotesize
$\displaystyle f(r_{in}-0) = f_0(r_{in}-0) + \tilde
f(r_{in}+\lambda) \frac{B_f(r_{in}-0)}{ B_f(r_{in}+\lambda)}$,}
{\footnotesize
$\displaystyle f(r_{ex}+0) = f_0(r_{ex}+0) + \tilde
f(r_{ex}-\lambda) \frac{B_f(r_{ex}+0)}{ B_f(r_{ex}-\lambda)}$,\\
}
\noindent where $B_f(r)$ is the amplitude function (envelope of
perturbations) and $\lambda$, the perturbation wavelength.
Subscript ``0'' indicates the equilibrium values. According to the
results of linear analysis (Paper I), the amplitude function at
the initial time instant is  $B_f(r) = r^{\beta_f}$. At
subsequent time instants $B_f$ can be computed by approximating
the minima and maxima of perturbations in the grid cells.
\end{itemize}

\begin{figure}
\caption{Radial dependences of relative perturbations (the solid bold, solid,
dashed, and dashed-and-dotted lines correspond to $\rho$, $U$, $p$, and $W$,
respectively) for $\theta = \theta_j$ at different time instants: $t=0.5$
(\textit{top}) and $t=1$ (\textit{bottom}). Large amplitudes correspond to
the  \mbox{$n_j=0$} harmonic of the waveguide-resonance mode and lower
amplitudes, to the Kelvin-Helmholtz mode. } \label{f1}
\end{figure}
\begin{figure}
\caption{Time dependences of the logarithms of density perturbation
amplitudes: the curves of families 1 and 2 correspond to $\theta = \theta_j$
and $\theta = 1.5 \theta_j$, respectively. The middle, lower, and upper
curves of each family on the upper panel correspond to $r(N_r/2)$, $r(3
N_r/4)$, and $r(N_r/4)$, respectively. The solid curves in the lower panel
were computed with a  $256 \times 128$ grid and the dashed curves, on a $128
\times 64$ grid.} \label{f2}
\end{figure}

\subsection{Parameter Values and the Transition to Dimensionless Variables}
\label{2-3}

The dimensionless parameter $\tau=t_{dyn}/t_{rad}$ that describes
the intensity of radiative cooling is determined by the ratio of
two time scales of the problem: the dynamic time scale $t_{dyn}$
and radiative cooling time scale $t_{rad}$. Here $t_{dyn} = r\,
\sin \theta_j/c_j$ is the time scale of the propagation of
perturbations from the boundary to the symmetry axis of the jet
($\theta=0$), and  $t_{rad} = p/[(\gamma-1)\,C_\Lambda \rho^2
\varepsilon^{5/2}]$ is the time during which the energy of gas
decreases by a factor of $e$ due to radiative cooling. At $\tau
\ll 1$ the effect of radiative cooling has a negligible effect on
the dynamics of perturbations (Paper I). At $\tau \ge 1$ radiative
cooling plays an important part in the evolution of unstable modes.

We do not take into account the processes connected to viscosity and heat
conductivity because the basic estimates show that the ratio
 of the
dynamical  time to the characteristic time of such processes has  the order $
{\mathrm {Kn} /\mathrm {M} \ll 1} $, where $ \mathrm {Kn} $ is the Knudsen
number and  $ M $ is the Mach number.

We performed simulations for $\gamma = 5/3$, $\tau = 2$ or $\tau =
4$, jet half-opening angles $\theta_j = 5^\circ$ or $\theta_j =
10^\circ$ and the jet-to-ambient medium pressure ratio of
$\tilde{R} = 5$ or $\tilde{R} = 10$.

In accordance with the results of Paper I, we set the Mach number
of the outflow equal to $M \simeq 0.69$ or $M \simeq 0.74$. For
the initial perturbations, we set $kr = 16$ and $\Delta_w =
10^{-9}$ or $\Delta_w = 10^{-8}$. We find no qualitative
differences between the results of the two series and therefore
below we describe the simulations with $\Delta_w = 10^{-8}$.
The results of  the series of the numerical simulations  have shown that
harmonics with $ {kr = 16} $ grow most quickly,  it is only these
harmonics that   we  show in the figures.

Note that we had to set such  a relatively large jet opening
angle because of the limited potential of the computers employed.
For perturbations to be processed correctly, the half-opening
angle must contain at least 10 cells of the computation domain,
whereas we could not increase  $N_\theta$ appreciably because of
the limited RAM of the computers employed.

Dimensionless boundaries of the computational domain were
determined by the $\tilde r_{in}=1$ and \mbox{$\tilde r_{ex}=e^\pi$}
values. In accordance with  Paper I, we perform
our analysis for the radii corresponding to the rigid-rotation
region of the galactic disk with the angular velocity
$\Omega \simeq 100 - 300\,km\,s^{-1}\mbox{kpc}^{-1} = {\rm const}$
(which corresponds to the region inside the bulge).

The outer boundary in dimensionless units is equal to $r_{ex}
\simeq (0.5 - 3)$ kpc and the inner boundary is $e^\pi$ times
smaller, $r_{in} \simeq (22 - 130)$ pc, depending on the specifics
of the rotation curves of real galaxies.

Concerning the transition to dimensionless velocities, let us point
out that the balance of pressures in the jet and in the ambient
medium (equation (4) in Paper I) allows equation (5) of Paper I
to be rewritten as:
\begin{equation}
\frac{1 }{ 2} \frac{\partial U_j^2 }{ \partial r} = \left(\frac{\rho_a}{
\rho_j} - 1 \right) \frac {\partial \Psi}{ \partial r}. \label{e7}
\end{equation}

We further take into account  that $U_j \propto r$ and the
dependence of gravitational potential on radius,  in
accordance with Paper I, is
\begin{equation}
\Psi = \Psi_0 + \frac{1 }{ 2} \Omega^2 r^2, \label{e8}
\end{equation}
to  find from  (\ref{e7}):
\begin{equation}
U_j = \Omega r \sqrt{\frac{\rho_a }{ \rho_j} - 1} = \Omega r \sqrt{\tilde{R}
- 1}. \label{e9}
\end{equation}

It is hence natural to normalize velocity to $\Omega r_{ex}$. In this case,
the sound speeds in the jet and in the ambient medium are unambiguously
determined by the dimensionless parameters $M$ and $\tilde{R}$. We have for
typical galaxies $\Omega r_{ex} \simeq (100 - 300)\,km\,s^{-1}$, and hence
(\ref{e9}) implies that we must be dealing with a high-velocity jet (for the
$\tilde{R}$ adopted above, e.g., $U_j = 3 \Omega r$) despite the subsonic
nature of the flow in this jet ($U_j < c_j$). This is consistent with the
conclusion that we made in Paper I that jet matter must be strongly heated
by the radiation of the galactic nucleus.

\begin{figure}
\caption{ Radial dependences of relative perturbations ($\rho$ (the solid
bold line); $U$ (the thin line); $p$ (the dashed line), and $W$ (the
dashed-and-dotted line)) at different time instants for $\theta =
\theta_j/2$:   $t=2$ ({\it top}), $t=3.2$ ({\it middle}), and $t=7.1$ ({\it
bottom}).} \label{f3}
\end{figure}
\begin{figure}
\caption{ Same as Fig.~\ref{f3}, but for $\theta = \theta_j$. } \label{f4}
\end{figure}
\begin{figure}
\includegraphics[scale=0.9]{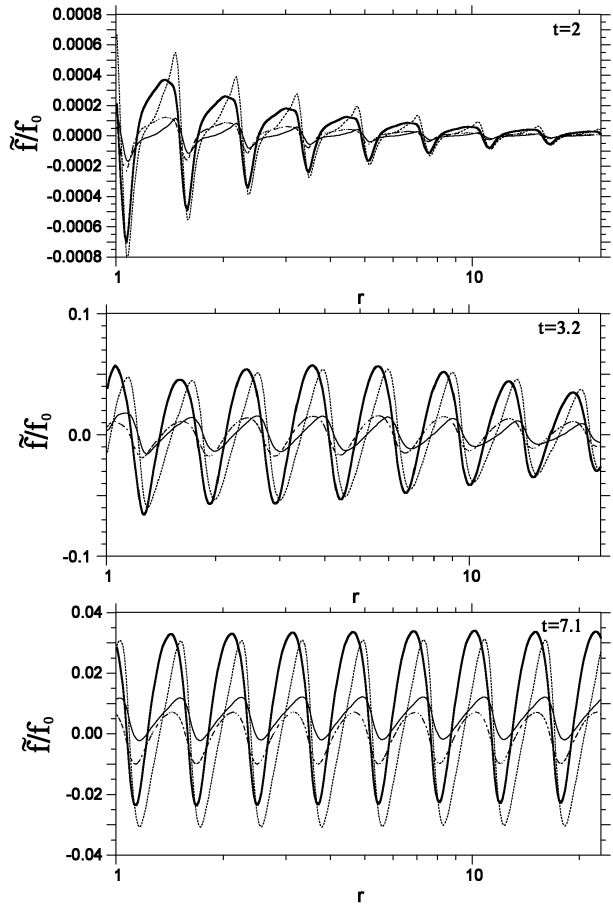}
\caption{ Same as Fig.~\ref{f3}, but for  $\theta = 1.25 \theta_j$. }
\label{f5}
\end{figure}
\begin{figure}
\includegraphics[scale=0.8]{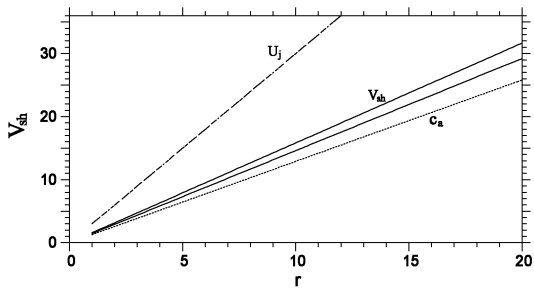}
\caption{Velocities of shock fronts at time $\tilde t =7.1$: angles $1.25
\theta_j$ (the upper solid line) and $1.5 \theta_j$ (the lower solid line).
The dashed and dashed-and-dotted lines show the sound speed in the ambient
medium  ($c_a$) and the jet velocity ($U_j$), respectively. } \label{f6}
\end{figure}

We determined the dimensionless time as $\tilde t = t
\Omega$. We have $\Omega =
(0.1-2)\times 10^{-14} \mbox{rad}\,\mbox{s}^{-1}$ and this corresponds to

 $t = (0.5- 9) \times 10^{14}\, \mbox{s} = (0.2-2.9)
\times 10^7$\,years.

Finally, to pass to dimensionless density, we chose the value
$\rho_j(r_{in})= 6.7 \cdot 10^{-23}\,\mbox{g}\,\mbox{cm}^{-3}$,
which corresponds to number densities $n_j(r_{in}) =
35\,\mbox{cm}^{-3}$.

\section{ RESULTS OF NONLINEAR NUMERICAL SIMULATIONS}
\label{3}

\subsection{Basic Comments}
\label{3-1}

When analyzing the evolution of perturbations, we could make use of higher
spatial resolution for axisymmetric modes (2D) and hence reproduce
higher-order effects. We therefore begin with the dynamics of pinch ($m =
0$) modes (Section \ref{3-2}), and only then discuss three-dimensional
helical  ($m = 1$) modes (Section \ref{3-3}).

Computation of the luminosity and surface density of the simulated
structures is associated with the following problems. It is evident from the
set of equations (\ref{e1})--(\ref{e5}) that the  real luminosity in
individual emission lines cannot be computed because the set includes no
equations of photoionization balance. However, structures discussed in the
following sections are observed inside the ionization cones both in Balmer
and shock-excited forbidden lines. Our analysis of the simulation results
yielded  the relative emission-line luminosity of the features of the wave
pattern by integrating function $(\rho^2 \Lambda(T) - \rho_0^2 \Lambda(T_0))
/ \rho_0^2 \Lambda(T_0)$ along the line of sight. In this definition one has
to subtract the contribution of hot matter inside the jet from the total
radiation. We therefore did not integrate the above function over the cells
located inside the jet ($\theta \le \theta_j$), because the ionized-gas
temperature at these sites is so high that the gas cannot radiate in optical
lines and most of its luminosity is contained at X-ray spectral domain.

\begin{figure*}
\includegraphics[scale=0.9]{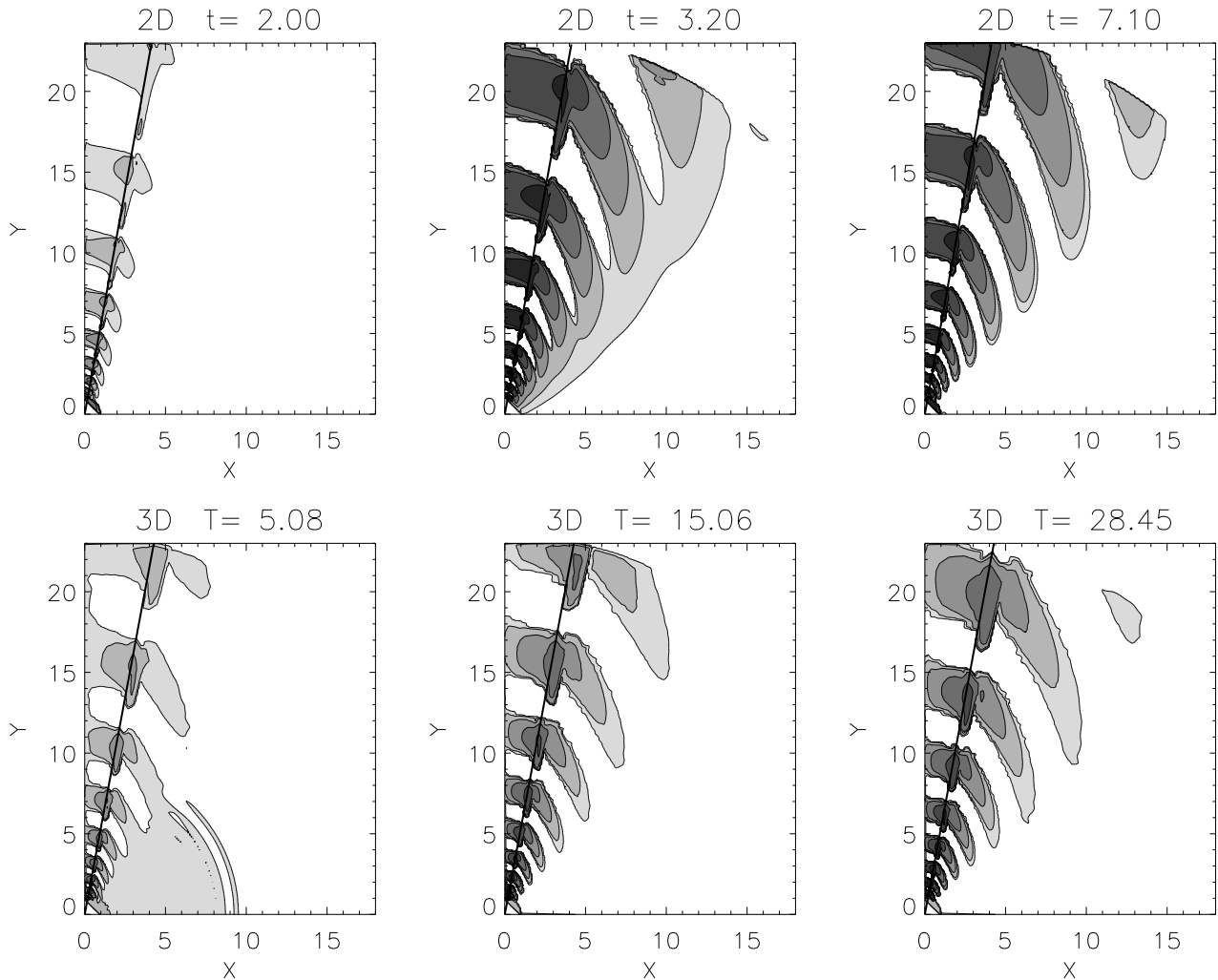}
\caption{ Luminosity contours for the two- (figures at the top) and
three-dimensional (figures at the bottom) simulations in the ($x=r\,\sin
\theta$, $y=r\,\cos \theta$) plane at different time instants.  The slanted
line shows the jet boundary ($\theta=\theta_j$). Darker shade indicates
higher luminosity values. } \label{f7}
\end{figure*}

\subsection{Results of 2D Simulation}
\label{3-2}

As we pointed out in Section \ref{2}, we performed a set of computations
with various parameter values and different numbers of cells in the
computational space: $N_r  =128$, $N_\theta=64,\,128,\,256$. Careful
comparison of eigenfunctions, obtained from the linear analysis, with
results of an initial stage of numerical simulations, shows that two
unstable modes developed in all the numerical simulations performed: the
surface Kelvin-Helmholtz (KHI) mode and the main\footnote{I.e., a mode with
no eigenfunction zero points between the jet boundary and symmetry axis ---
$n_j = 0$ (see Paper I).} waveguide-resonance IGW mode of the $u^-$ family.
Unlike the latter, the Kelvin-Helmholtz mode proved to be very sensitive to
the thickness of the transition layer between the jet and ambient medium,
which is determined by the step of the grid cell size in the
$\theta$--direction. The stabilizing effect of the velocity shift is a well
known fact (see, e.g., \cite{bodo}; \cite{musk}). Therefore the KHI mode
contributed appreciably to the resulting perturbation pattern only on the
$N_\theta = 256$ grid (see Fig.~\ref{f1}), whereas its amplitude was
negligible in all other computations.

Figure~\ref{f2} illustrates the above-mentioned effect of the ``thickness"
of the transitional layer on the perturbation increment: increment grows
with increasing spatial resolution of the grid and asymptotically tends to
the value determined from linear analysis of the discontinuous model.

A characteristic feature of the evolution of perturbations in all the
computational series performed is that radial dependences of the amplitude
envelopes of all quantities can be fitted fairly well by function
$r^{\beta_f}$ (see Section 3 of Paper I). At the same time, at the stage of
nonlinear saturation these dependences transform to the form $r^{\alpha_f}$,
which describes equilibrium distributions (see Figs.~\ref{f3}--\ref{f5}).

During the nonlinear stage the development of waveguide-resonance
instability of the main ($n_j = 0$) IGWs mode produces shocks in the medium
that surrounds the jet with approximately paraboloidal fronts in the real
three-dimensional space. Figure~\ref{f5} shows the radial dependence of the
$r$ component of mass velocity at $\theta = 12.5^\circ$. The perturbations
discussed here can be seen to be real shocks given the well-known fact that
the discontinuity surfaces in computations based on the TVD-E numerical
scheme are always ``smeared'' over at least five spatial cells of the system
(\cite{ryu}). It is clear from Fig.~\ref{f5} that the width of the shock in
our case is equal to four to six cells, i.e., we indeed observe a
discontinuity surface --- a shock front --- at this location. A comparison
of the perturbation propagation velocity $v_{sh}$ in the direction normal to
the shock surface with the sound speed $c_a$ in the medium that surrounds
the jet corroborates this conclusion. As is evident from Fig.~\ref{f6}, the
shock-front velocity is supersonic and our analysis shows that the Mach
number  of the shock front, $M_{sh} = v_{sh} / c_a$, does not change with
radius. We believe that this increase of the velocity of the shock fronts
with increasing radius can be explained by the combined effect of two
factors. First, the development of instability with the distance from the
center. Second, the well-known effect of the acceleration of the shock as it
propagates toward decreasing density, because the distribution of the
gravitational potential (\ref{e8}) adopted in our model implies that $\rho
\propto r^{-3}$.

Figure~\ref{f7} demonstrates the evolution of the relative perturbations of
luminosity normalized to its equilibrium value. The propagation of shocks
from the jet boundary into the ambient medium is immediately apparent.

Let us dwell on the following point, which is of great importance for our
analysis.  A characteristic feature of the systems with Newtonian-type
gravitational potentials (protostars) is that perturbations born in the jet
encompass the entire surrounding ``atmosphere''. This is due to impedance in
the atmosphere being lower than in the jet: $\rho_a c_a < \rho_j c_j$. That
is why we can assume that in a protostar-type accretion-jet system the jet
and the accretion disk interact through waves in the atmosphere
(\cite{levin}).  In our model, we have a reverse situation: $\rho_a c_a >
\rho_j c_j$, which prevents the penetration of perturbations into the
ambient medium. In this case, shocks propagate inside a limited cone with
the half-opening angle of $\theta_{cone}$ about the central jet.

\begin{figure}[!h]
\includegraphics[width=8.0cm]{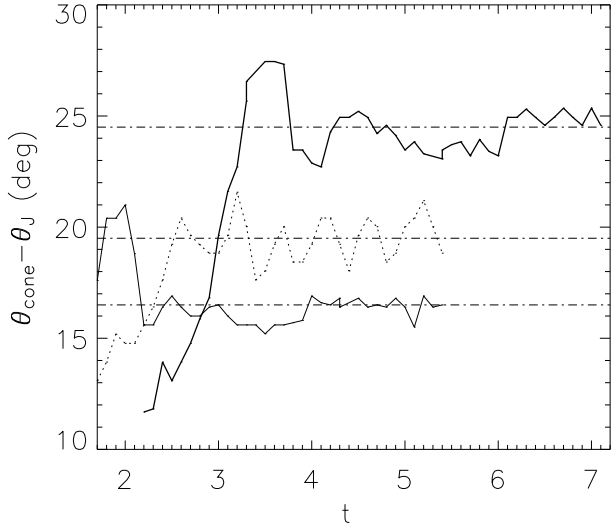}
\caption{Time dependence of the opening angle of the conical domain
overtaken by perturbations. The bold solid line shows the results of the
simulations with a $256 \times 128$ grid, $\theta_j = 10^\circ$, $\tau = 2$,
and $\tilde{R} = 10$. The thin solid line shows the results of the
simulations with a $256\times256$ grid, $\theta_j = 5^\circ$, $\tau = 4$, and
$\tilde{R} = 10$. The dashed line shows the results of the simulations with
a $256 \times 256$ grid, $\theta_j  = 5^\circ$, $\tau = 4$, $\tilde{R} = 5$.
} \label{f8}
\end{figure}

Figure~\ref{f8} shows the dependence of the characteristic depth of such a
penetration in latitude $\theta$ \mbox{($\theta_{cone}-\theta_j$)} as a
function of time. As is evident from this figure, $\theta_{cone}$ first
increases rapidly with time and then reaches saturation. Note that this
effect --- localization of perturbations in the ambient medium inside a wide
cone around the jet --- does not depend on the mode type, because impedance
is independent of the particular type of wave symmetry (symmetric or
helical). Moreover, we argue that this effect is also independent of the
physical mechanism of mode excitation. This means that perturbations of both
volume resonance modes due to superreflection and Kelvin--Helmholtz surface
modes decrease equally rapidly in the ambient medium with the distance from
the jet boundary.

\begin{figure}[tbp]
\includegraphics[scale=0.8]{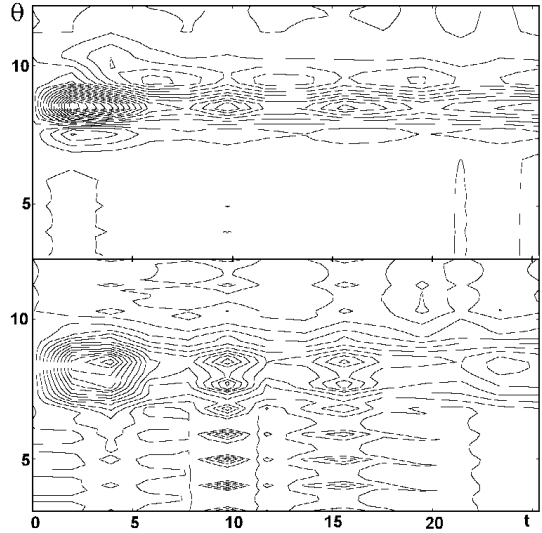}
\caption{Contours of radially-averaged distributions of temperature $(T(t) -
T(0))   / T(0)$ (top) and the $\theta$--component of velocity $W(t)$
(bottom).} \label{f9}
\end{figure}

As is evident from Fig.~\ref{f8}, the opening angle of the cone overtaken by
perturbations exhibits quasi-periodic damped oscillations. This behavior is
due to the above-mentioned nonlinear superposition of two modes with
different physical buildup mechanisms. These modes have different
frequencies and, correspondingly, beating shows up at the half-difference
frequency superimposed on the carrier frequency equal to the half-sum of the
mode frequencies. After reaching nonlinear saturation, the KHI begins to
smoothly decay and so does beating, and subsequent evolution is determined
by the dynamics of the volume resonance mode $n_j = 0$. Quasi-periodic
oscillations show up most conspicuously in the contours of radially-averaged
perturbations drawn in the ``latitude --- time'' plane (see Fig.~\ref{f9}).

\begin{figure}
\includegraphics[width=8.5cm]{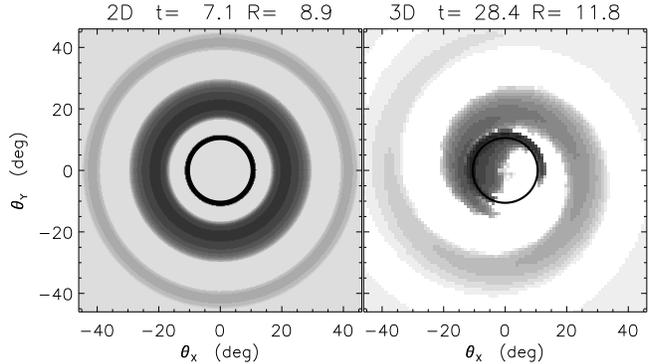}
\caption{Cross sections of the volume-density cone at the constant-radius
level. Dark shades correspond to maximum-brightness regions. Left-hand panel
shows the axisymmetric mode in the 2D simulation (the luminosity cone
obtained by rotating the computed domain about the $\theta=0$ axis). The
right-hand panel shows the first helical mode according to the results of 3D
simulations. The central circle corresponds to the jet boundary. The
corresponding radius and time are indicated above each figure. } \label{f10}
\end{figure}

Oscillations are due to yet another effect. Heating of gas in
shocks changes the dispersion law. This results in the decrease
of the flux of energy transferred to perturbations due to the
development of instability, and the decrease of the perturbation
amplitude. Subsequent radiative cooling causes the relaxation of
gas to a close-to-equilibrium state, the dispersion law is
restored, and the perturbation amplitude grows again. Both
processes described above (beating and quasi-periodic variations
of the dispersion law) are essentially nonlinear and difficult to
decouple from each other.

\subsection{Results of 3D Simulations}
\label{3-3}

In the case of 3D simulations we are forced to restrict our analysis to a
$128 \times 64 \times 32$ grid, which is rather coarse in terms of the
$\theta$ and $\varphi$ angles. Therefore the growth of perturbations due to
the effect described in Section \ref{3-2} was much slower than in the case
of 2D simulations and ceased at lower relative amplitudes. It is, however,
safe to say that all the basic patterns of the evolution of perturbations are
the same as in the two-dimensional case and therefore here we discuss only
the aspects that differ.

\begin{figure*}[tbp]
\includegraphics[scale=0.9]{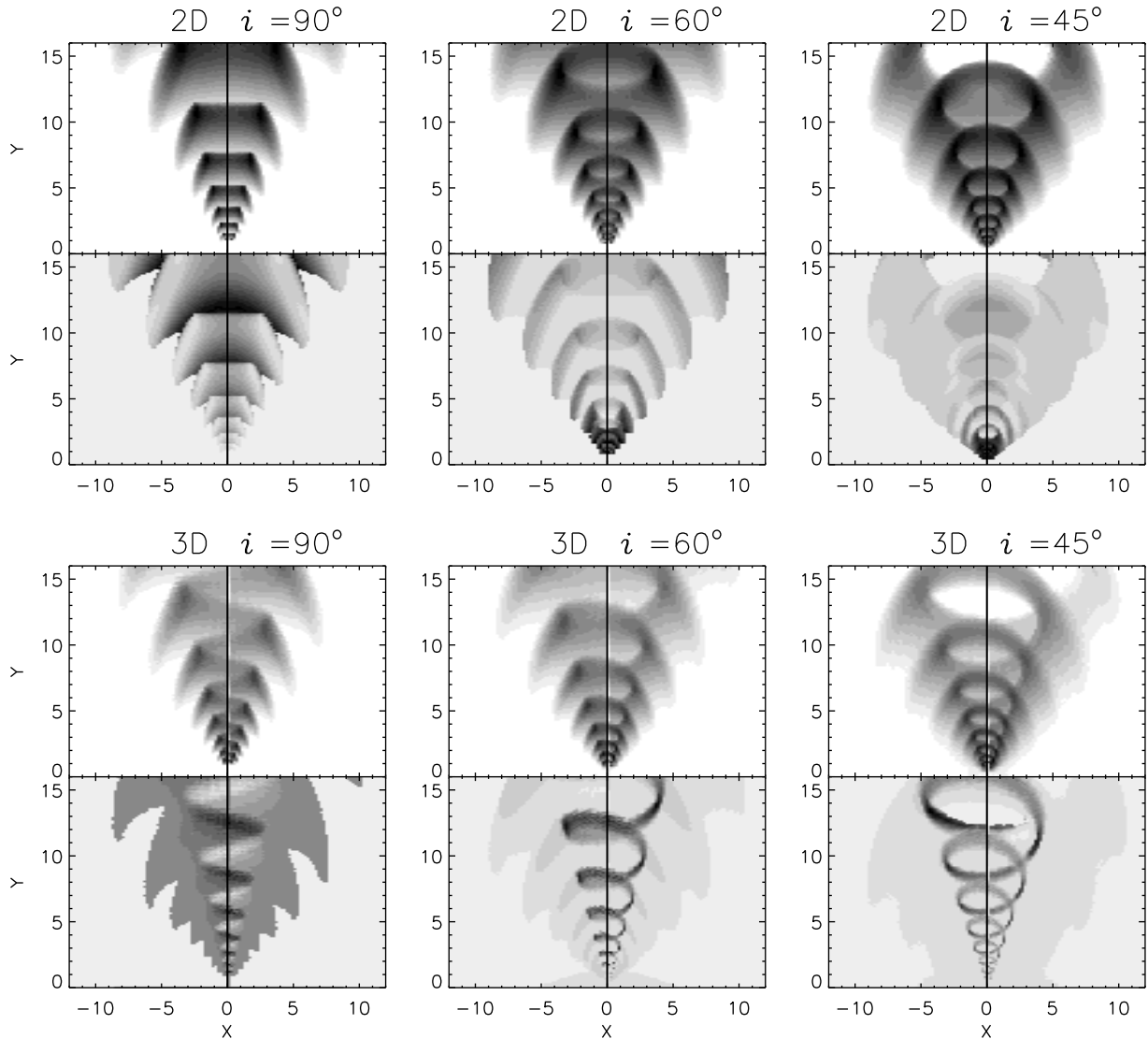}
\caption{ Sky-plane projection of emission-line cones for different tilt
angles $i$ between the line of sight and cone axis. The upper block of
figures shows the results of 2D simulations (axisymmetric perturbations).
The lower block of figures shows the results of 3D simulations (the first
helical mode of perturbations). The upper maps show the luminosity (surface
density) maps and the lower maps correspond to radial-velocity fields in
arbitrary scale.} \label{f11}
\end{figure*}

First, in the case of 3D simulations we analyzed the development
of the first helical ($m = 1$) mode until the development of
shocks. In the constant-$r$ cross section the wave pattern of
this mode has the form of a one-armed trailing (relative to the
rotation of the phase pattern) spiral
--- see Fig.~\ref{f10}. In three dimensions this pattern has the form of a
slowly rotating (compared to $\Omega r$) helical structure filling
a wide cone around the central collimated jet.

We constructed a series of simulated wave-pattern surface brightness maps
(``emission-line images'', see Section~ \ref{3-1}) for different tilt angles
$i$ of the cone axis to the line of sight. We also constructed the velocity
fields, i.e., the maps of luminosity-weighted line-of-sight gas velocities.
Figure~\ref{f11} shows examples of such maps separately for the symmetric
and helical patterns.

It is evident from the figures that the sky-plane projections of
the three-dimensional helical pattern depend on angle $i$,
allowing both the $Z$--shaped and loop-shaped emission-line
patterns to be obtained. The rotation of this phase feature on the
radial-velocity map results in switching from ``red'' to ``blue''
shift to be observed as we pass from one arm of the $Z$--shaped
pattern to another (Fig.~12). We pointed out in  the
``Introduction'' to Paper~I that such a velocity behavior is
typical of some galaxies with $Z$-shaped emission-line features.
In our model switch of velocities is explained first and foremost
by the rotation of the wave pattern inside the cone.

\subsection{Astrophysical Applications}
\label{3-4}

Thus our simulations show that instabilities of highly collimated
jets emerging from active galactic nuclei result in the
propagation of shocks from the jet boundary into the ambient
medium. These shock waves are located in the cone with a
half-opening angle of $15^\circ \div 40^\circ$ outside the jet. It
is evident that in actual situations the ionized gas behind the
shock should radiate intensively in optical lines ($H_\alpha$,
[OIII] etc.). A rather short-wavelength structure forms
--- $kr \simeq 15 - 20$ --- and hence even a relatively small
tilt of the symmetry axis of the cone considered with respect to
the observer would produce a projection effect making the
structure to appear as a continuous radiation cone with brighter
regular features --- ``arcs'' for axisymmetric wave modes or a
Z-shaped pattern for the first helical mode --- superimposed on
it.

In Fig.~\ref{f11} we sketchy reproduce the situation described based on the
results of our simulations for three different tilt angles of the cone
symmetry axis to the line of sight and show the distribution of luminosity
integrated along the line of sight (thus the adopted model assumes optical
transparency of gas). It goes without saying that in real galaxies one has
to take into account dust absorption behind the shock fronts. Thus the
images of ionization cones reported by Ferruit et al. (1998) and Quillen et
al. (1999)  exhibit extended dust lanes associated with regular structures
inside the cones.

The proposed scenario of the formation of ionized cones makes it also
possible to describe to a first approximation the radial-velocity pattern
observed in these cones. In the previous section we already explained
velocity switching inside a $Z$-shaped emission-line structure due to a
rotating helical wave. At the same time, the velocity fields of some Seyfert
galaxies exhibit only ``blue'' Doppler shifts in one ionization cone and only
``red'' shifts in the diametrically opposite cone. A good example is
provided by the NGC~5252 galaxy whose velocity field was published by Morse
et al. (1998) and Moiseev  et al. (2000). Below we show that such a behavior
can be explained by the development of an axisymmetric wave mode.

\begin{figure}[t]
\includegraphics[width=8.50cm]{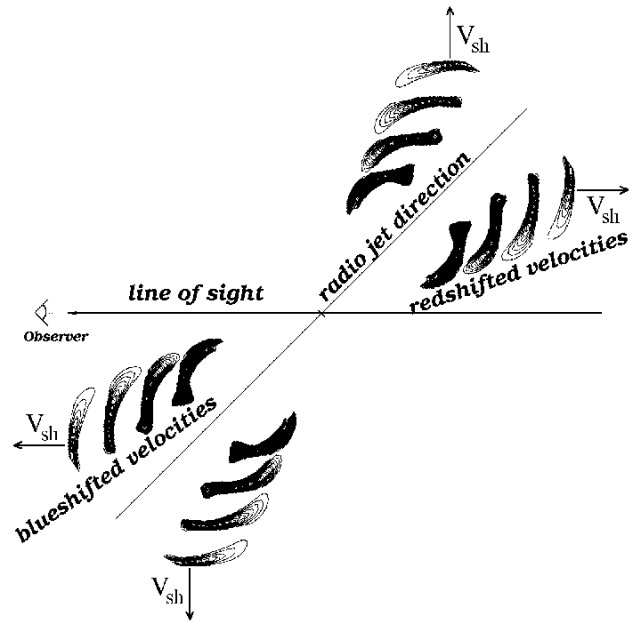}
\caption{A diagram explaining the formation mechanism of the observed
velocity field in the case of the development of an unstable axisymmetric
mode. We adopted the shock fronts in the $r - \theta$ plane from the results
of our 2D simulations for dimensionless time instant $t = 10$.} \label{f12}
\end{figure}

Indeed if the tilt angle of the radio-jet axis to the line of sight  $\alpha
\simeq 30-60^\circ$, then the cone nearest to the observer should be seen
through shock fronts propagating virtually toward the observer and the
farthest cone, against receding shock fronts (see Fig.~\ref{f12}). It is
evident from Fig.~\ref{f6} that the velocities of these fronts can be
estimated as:
\begin{equation}
\begin{array}{l}
{\displaystyle v_{sh} \simeq (1.2 \div 1.6) c_a = (1.2 \div 1.6) {U_j \over
\sqrt{\tilde{R}} M} = }\\ \qquad \qquad \qquad \qquad \qquad {\displaystyle =
(1.2 \div 1.6) {3 \Omega r \over \sqrt{\tilde{R}} M}. }
\end{array}
 \label{e10}
\end{equation}
We now substitute the numerical values quoted in~\ref{2-3} into (\ref{e10})
to determine $$v_{sh}\simeq(150-630)\,km\,s^{-1},$$ which agrees well with
the observed Doppler shifts (see references to Paper~I). Moreover, the
velocities of shock fronts increase linearly with radius (see Fig.~\ref{f6}),
 implying that the
Doppler shift of radial velocities should also increase with the
distance from the nucleus.

\begin{figure*}

\includegraphics{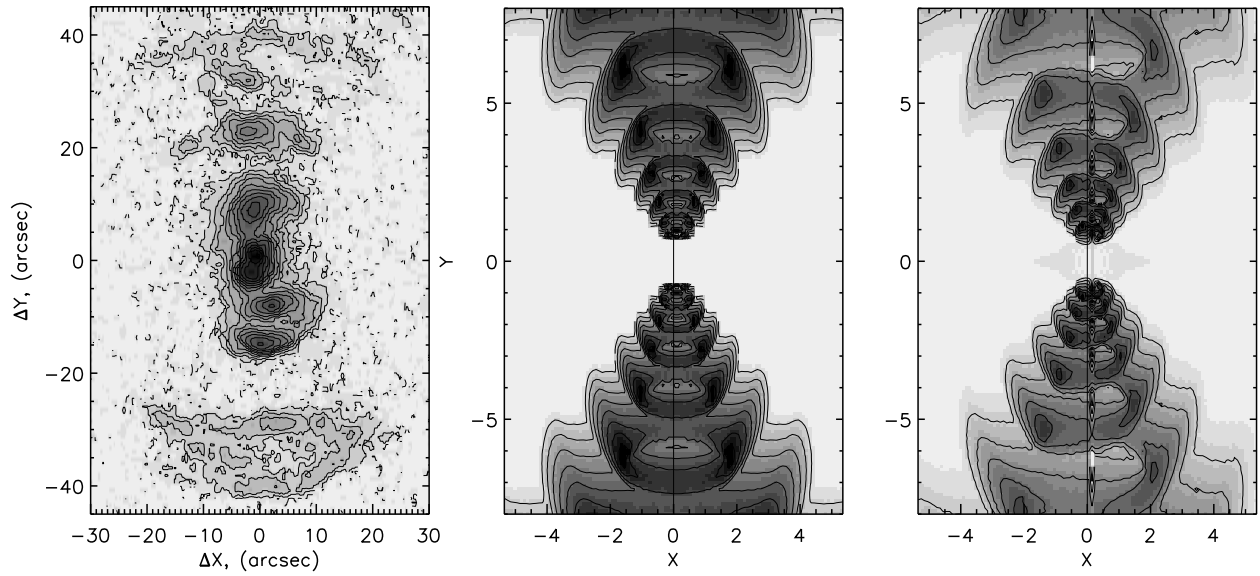}
\caption{ Left-hand panel: an [OIII]-line contour image of the NGC 5252
galaxy. Central panel: simulated velocity map for the pinch mode  ($m=0$)
$i=60^\circ$. Right-hand panel: simulated luminosity map for the helical
mode ($m=1$) $i=60^\circ$. } \label{f13}
\end{figure*}

In Fig.~\ref{f13} we compare the [OIII]-line image of the NGC~5252 galaxy
(contour image in Fig.~1 from Paper~I) and examples of arbitrarily scaled
simulated images for the modes with $m=0$ and $m=1$.

Note that our simulations were either axisymmetric or
nonaxisymmetric. However, in real situations pinch perturbations
and  helical modes in the jet develop simultaneously. Nonlinear
superposition of such modes can explain the morphology of the
wave pattern and the velocity field observed in the NGC 5252
galaxy if we assume that the helical and axisymmetric (pinch)
mode dominate in the inner and outer parts, respectively.

\section{DISCUSSION AND CONCLUSIONS}
\label{4}

\begin{itemize}
\item  Results of our nonlinear numerical simulations qualitatively confirm the
conclusions based on linear analysis made in Paper~I.

\item  The development of the waveguide-resonance instability of
internal gravity modes in a highly collimated jet outflow from
the galactic nucleus may result (in the domain controlled by the
gravity of the bulge) in the formation of a short-wavelength ($kr
\simeq 15-20$) periodic system of shocks (Mach cones) in the
matter surrounding the jet.

\item  The shock structure mentioned above embraces a wide cone
with a half-opening angle of $\theta_c \simeq 15^\circ-40^\circ$.
The opening angle of the cone is determined by particular
parameters of the outflow (first and foremost, by the ratio of
the impedances of the jet and the ambient gas).

\item Given the intense radiative cooling of gas behind the shock
fronts and projection effects, it can be assumed that the shock
system discussed here must appear to the observer as a wide
radiation cone with a wave pattern superimposed. In this case,
helical modes determine the appearance of Z-shaped emission-line
structures observed in a number of Seyfert galaxies.

\end{itemize}

Note that a number of authors already solved a similarly formulated problem
(see, e.g., \cite{rossi}; \cite{limsteffen}; \cite{hardee}), however, they
analyzed an earlier stage of the interaction between the jet and the
surrounding gaseous clouds. A ``cocoon" does indeed form behind the bow
shock when a supersonic jet intrudes into the ambient medium. This effect was
demonstrated many times in numerical simulations and one can see it on
images made with the Hubble Space Telescope, e.g., in the central region of
Mrk~3 (\cite{cap99}) or Mrk~78 (\cite{whittle}). However, on longer spatial
scale lengths ($r>0.5 - 1$ kpc), like in the case of NGC\,3516, Mrk\,573,
and NGC\,5252 (see images in Paper\,I), no such shock is observed. This
unambiguously proves that  the initial injection of the jet (which formed
deep in the inner region of the system in the accretion torus around the
black hole) occurred rather long ago. We believe that in these objects the
jet has already punched a channel in a more or less dense medium in the
central region of the galactic disk, and the bow shock has broken into the
intergalactic medium (like this happens with galactic fountains). Since
then, the gas of the cocoon must have relaxed (at least partially) and it
must interact with the jet over the scale lengths mentioned above, leading
to instability. On long scale lengths the jet propagates ballistically,
however, its gas loses energy as a result of radiative cooling and there are
no new mechanisms for the buildup of oscillations (because of the extremely
low density of the ambient medium), implying that gas does not show itself
in any observable way. We simulated exactly this situation. Note the
instability of this type may develop even in the absence of a sharp jet
boundary. It is sufficient for the supersonic velocity difference to occur
over less than one wavelength, and there will always be such perturbations.

We understand that our model is only a first step toward the
construction of the pattern of regular structures in ionization
cones. When constructing a self-consistent model of such objects
one must take into account at least the following factors:

\begin{itemize}
\item  Contributions of the flat (large-scale disk) and triaxial (bar)
subsystems.

\item Rotation of the galactic nucleus.

\item  Slow rotation of the matter of the conical jet about its symmetry
axis.

\item  Interaction between the jet and the ambient
gas of the disk (in the case of a small angle between the
jet and the disk plane.

\item The presence inside the shock of a ``cocoon''
produced by the supersonic intrusion of gas of the
conical outflow into the dense gas of the disk.
\end{itemize}

We believe the above factors to be of great importance for the
dynamics of various concrete objects, however, they do not
determine the essence of the phenomenon and therefore we ignored
them in this paper. However, when discussing the formation of the
observed structures in concrete objects one has to take most of
the above factors into account.

In addition, to conclude our discussion, we point out that
collimated jet outflows from active galactic nuclei leave the
bulge region after having acquired an essentially nonlinear, shock
structure. Let us also add that near the bulge region (in the
sense that the gravitational potential is dominated by the
spherically symmetric part) this structure must undergo
significant changes due to rather sharp change of the law of the
distribution of the gravitational potential and hence that of
pressure and density inside the jet and in the ambient medium.

\begin{acknowledgements}
We are grateful to V.V.~Levi and A.V.Nikitin for their assistance
and numerous useful discussions. A.V.Moiseev and V.L. Afanasiev
acknowledge the support of the Russian Foundation for Basic Research  (project
no.~06-02-16825).
We thank the reviewer Torgashin Yu.M. for valuable remarks that improved
this paper.
\end{acknowledgements}

\end{document}